\documentclass[aps,pra,onecolumn,showpacs,floatfix,
superscriptaddress,reprint]{revtex4-1}
\usepackage[T1]{fontenc}
\usepackage{graphicx}
\usepackage{bm}
\usepackage{color}
\usepackage{txfonts}
\usepackage{times}
\usepackage{multirow}
\usepackage{soul}
\usepackage[colorlinks=true, linkcolor=blue,
citecolor=blue, urlcolor=blue]{hyperref}

\definecolor{cred}{RGB}{180,0,0}

\begin{document}

\newcommand{\todo}[1]{{\color{cblue}\textbf{[#1]}}}

\title{Mass scaling and non-adiabatic effects in photoassociation spectroscopy
of ultracold strontium atoms}

\author{Mateusz Borkowski}
\author{Piotr Morzy\'nski}
\author{Roman Ciury\l o}

\affiliation{Institute of Physics, Faculty of Physics, Astronomy
	and Informatics, Nicolaus Copernicus University,
	Grudziadzka 5, 87-100 Torun, Poland}

\author{Paul~S.~Julienne}

\affiliation{Joint Quantum Institute, University of Maryland and National Institute of Standards and Technology, College Park, Maryland 20742, USA}

\author{Mi Yan}

\author{Brian J. DeSalvo}

\author{T. C. Killian}

\affiliation{Rice University, Department of Physics and Astronomy,
  Houston, Texas, 77251}

\date{\today}

\begin{abstract} 
We report photoassociation spectroscopy of ultracold $^{86}$Sr atoms near the
intercombination line and provide theoretical models to describe the obtained
bound state energies.  We show that using only the molecular states correlating
with the $^1S_0$$+$$^3P_1$ asymptote is insufficient to provide a mass scaled
theoretical model that would reproduce the bound state energies for all isotopes
investigated to date: $^{84}$Sr, $^{86}$Sr and $^{88}$Sr. We attribute that to
the recently discovered avoided crossing between the $^1S_0$$+$$^3P_1$ $0_u^+$
($^3\Pi_u$) and $^1S_0$$+$$^1D_2$ $0_u^+$ ($^1\Sigma^+_u$) potential curves at
short range and we build a mass scaled interaction model that quantitatively
reproduces the available $0_u^+$ and $1_u$ bound state energies for the three
stable bosonic isotopes. We also provide isotope-specific two-channel models
that incorporate the rotational (Coriolis) mixing between the $0_u^+$ and $1_u$
curves which, while not mass scaled, are capable of quantitatively describing
the vibrational splittings observed in experiment. We find that the use of
state-of-the-art \emph{ab initio} potential curves significantly improves the
quantitative description of the Coriolis mixing between the two $-8$~GHz bound
states in $^{88}$Sr over the previously used model potentials. We show that one
of the recently reported energy levels in $^{84}$Sr does not follow the long
range bound state series and theorize on the possible causes.  Finally, we give
the Coriolis mixing angles and linear Zeeman coefficients for all of the
photoassociation lines. The long range van der Waals coefficients
$C_6(0_u^+)=3868(50)$~a.u. and $C_6(1_u)=4085(50)$~a.u. are reported.
\end{abstract}

\maketitle

\section{Introduction}

\parskip 0pt

Photoassociation (PA) spectroscopy is a widely used tool for the study of atomic
collisions and determination of bound state energies of diatomic molecules
\cite{Jones2006}. In this process two colliding cold atoms are bound together
into an excited molecule by optical excitation. A recent focus for PA
spectroscopy has been the study of molecules created by excitation to the red of
an intercombination line ($^1S_0$$\rightarrow$$^3P_1$) in divalent atoms such as
alkaline-earth metal atoms calcium \cite{Kahmann2014} and strontium
\cite{Zelevinsky2006, McGuyer2013, Stellmer2012}, and the rare-earth atom
ytterbium \cite{Tojo2006, Borkowski2009, Borkowski2011, Takasu2012}. The
interest in intercombination line PA spectroscopy \cite{Ciurylo2004} is driven
by its potential applications in the production of ground state ultracold
molecules \cite{Skomorowski2012a}, the potential for control of scattering
lengths in ultracold collisions via optical Feshbach resonances
\cite{Ciurylo2005}, as well as coherent photoassociation \cite{Koch2012, Mi2013}
and finally electron-proton mass ratio measurements \cite{Kotochigova2009}.

The interactions between strontium atoms, both in their ground and excited
states, have been extensively studied by means of intercombination line PA
spectroscopy. Zelevinsky~\emph{et al.} \cite{Zelevinsky2006} made the first
observations with $^{88}$Sr atoms confined in an optical lattice and recently
Stellmer \emph{et al.} \cite{Stellmer2012} have reported similar measurements with
$^{84}$Sr atoms. Two-color photoassociation spectroscopy enabled Martinez
de~Escobar~\emph{et al.} \cite{MartinezdeEscobar2008} to accurately determine
the scattering lengths of all Sr isotopes which helped to explain the low
thermalization rate in $^{88}$Sr. In addition to the PA spectroscopy studies, both the
interactions in the ground \cite{Stein2008, Stein2010} and excited
\cite{Stein2011} states of the Sr$_2$ molecule have been studied by Fourier
transform spectroscopy.

The research into intercombination line PA is fueled by the possible use of
optical Feshbach resonances (OFRs) \cite{Fedichev1996, Bohn1997, Chin2010} to
enable optical control of the scattering lengths. This is especially important
for ground state strontium atoms, because its $^1S_0$ ground state and a lack of
hyperfine structure in its bosonic isotopes precludes the existence of magnetic
Feshbach resonances in this system.  Early experiments with Na
\cite{Lett1993,Fatemi2000} and Rb \cite{Theis2004, Thalhammer2005} have shown
that the usefulness of OFRs in alkali-metals is greatly hindered by the large
loss of atoms due to photoassociation. However, in the case of the narrow
intercombination lines in divalent atoms like Ca, Sr, and Yb, these losses can
be greatly reduced \cite{Ciurylo2004, Ciurylo2005} and useful changes in
scattering lengths have been shown for both Yb \cite{Enomoto2008} and recently
Sr: a proof-of-concept investigation in a thermal gas \cite{Blatt2011} and an
example of the use of OFRs as a means of controlling the collapse of an Sr
Bose-Einstein condensate (BEC) \cite{Yan2013}.

All isotopes of Sr have been brought to quantum degeneracy. The most abundant
isotope $^{88}$Sr is known for its small negative scattering length
\cite{Mickelson2005, MartinezdeEscobar2008}, which thwarted the early attempts
\cite{Ido2000, Ferrari2006} at quantum degeneracy. The least abundant isotope,
$^{84}$Sr, has excellent collision properties for evaporative cooling, so it was
the first Bose-condensed isotope \cite{Stellmer2009, MartinezdeEscobar2009}.
Since then, the isotope $^{86}$Sr has also been condensed \cite{Stellmer2010},
while the thermalization problem in $^{88}$Sr has been circumvented by
sympathetic cooling with $^{87}$Sr \cite{Mickelson2010}. The narrow
intercombination line enabled direct laser cooling of $^{84}$Sr down to quantum
degeneracy \cite{Stellmer2013b}. Degenerate Fermi \cite{DeSalvo2010} and
Bose-Fermi gases \cite{Stellmer2013a} have also been reported. Strontium is being
actively explored for its use in the making of ultracold molecules: ground state
strontium dimers \cite{Skomorowski2012a, Reinaudi2012}, and the heteronuclear
RbSr molecules \cite{Tomza2011}. Rubidium-strontium mixtures have become
especially promising after a degenerate quantum mixture of Rb and Sr atoms was
obtained \cite{Pasquiou2013}.

We report energy levels of the $^1$S$_0$+$^3$P$_1$ Sr$_2$ molecule obtained for
the $^{86}$Sr isotope, which complements the currently available data for
$^{88}$Sr \cite{Zelevinsky2006} and $^{84}$Sr \cite{Stellmer2012}, and we
provide theoretical models of the interactions in the Sr$_2$ molecule. A set of
energy levels in the subradiant $1_g$ state of the strontium dimer has also
become available \cite{Zelevinsky2014}, but is outside the scope of this
article. The paper is organized as follows. In Section \ref{sec:experiment} we
briefly describe the experimental details and the PA data obtained for
$^{86}$Sr. In Section \ref{sec:longrange} we provide a theoretical model based
on recent state-of-the-art \emph{ab initio} potential curves
\cite{Skomorowski2012} for the description of the long range interactions in
this excited state of the strontium dimer. We will use this model in Section
\ref{sec:vibrational} to provide a quantitative description of vibrational
splittings,   as well as the nonadiabatic Coriolis effects and linear Zeeman
coefficients \cite{McGuyer2013} for all photoassociation lines reported to date.
In the case of one of the isotopes, $^{84}$Sr, we will find a significant
discrepancy between one of the experimental \cite{Stellmer2012} and theoretical
positions of one of the lines and theorize on its possible causes. In the case
of $^{88}$Sr we will show that the use of realistic potential curves
significantly improves the quantitative description of the positions of strongly
Coriolis-mixed energy levels over previous work \cite{Zelevinsky2006}. We also
show that the improvement of the description of Coriolis mixing is followed by
better agreement of the respective Zeeman g-factors with the experimental data
\cite{McGuyer2013}. Finally, in Section \ref{sec:massscaling} we will
investigate the mass scaling between strontium isotopes, that is, the
possibility of using the same potential curves for the description of PA
spectroscopy data for all isotopes. While mass scaling of only the long range
potentials was sufficient in the description of the energy levels near the
$^1$S$_0$+$^3$P$_1$ asymptote in different isotopes of a similar species
ytterbium \cite{Borkowski2009}, it fails in the case of strontium. We will
explain this effect for the $0_u^+$ series quantitatively by augmenting our
model with the recently discovered \cite{Stein2011, Skomorowski2012} curve
crossing between the $^1$S$_0$+$^3$P$_1$ $^3\Pi_u$ curve (which supports the
$0_u^+$ series) and a $^1\Sigma_u^+$ curve correlating, remarkably, to the
$^1$S$_0$+$^1$D$_2$ asymptote. Once mass-scaling of the $0_u^+$ bound states is
achieved we add a third channel representing the $^1S_0$+$^3P_1$ $1_u$ state.
The final three-channel mass scaled model reproduces the available $0_u^+$ and
$1_u$ bound state energies to within $0.5$ MHz on average.

\section{Photoassociation spectroscopy of $^{86}$Sr \label{sec:experiment}}

\begin{table*}[htbp]
 \caption{Comparison of available experimental $^1S_0$$\rightarrow$$^3P_1$
 photoassociation data with our long range interaction model. The experimental
 data encompasses all bosonic isotopes: $^{88}$Sr (for total angular momentum
 $J=1$ \cite{Zelevinsky2006}, and recently $J=3$ \cite{McGuyer2013}), $^{84}$Sr
 data from Stellmer \emph{et al.} \cite{Stellmer2012}, and the previously
 unreported data for $^{86}$Sr. The numbers in parentheses denote
 the experimental uncertainty, where available. The model parameters were
 initially fitted to the bound state energies for $^{88}$Sr and then extended to
 other isotopes by only changing the respective short range
 wavefunction phase. For $^{88}$Sr we also cite the theoretical bound state
 energies from the original Zelevinsky \emph{et al.} \cite{Zelevinsky2006} model
 to show the improved description of Coriolis mixing between the two bound
 states at -8430 and -8200 MHz that our model provides.  For
 $^{84}$Sr we find it difficult to fit all of the energy levels together,
 therefore the results shown are based on a fit that omits the $-1288$ MHz
 state. The observed drastic discrepancy between this state and the theoretical
 energy suggests that the $-1288$ MHz may not be a member of the $0_u^+$ $J=1$
 series. This is discussed in Section \ref{sec:vibrational}. We also give the
 Coriolis mixing angles and Zeeman g-factors as measured in Ref.
 \cite{McGuyer2013} and their theoretical counterparts as described in Section
 \ref{sec:zeeman}.}
\begin{ruledtabular}
    \begin{tabular}{rrrrrrrrrrr}
    & & & \multicolumn{3}{c}{Binding energy $E/h$ (MHz)} & \multicolumn{2}{c}{Mixing angle $\theta$} & \multicolumn{3}{c}{Linear Zeeman coefficient $g$ } \\
    \cline{4-6} \cline{7-8} \cline{9-11} \noalign{\vskip 2mm}
 Isotope & Series & $J$ & Exp. \cite{Zelevinsky2006} & Theory \cite{Zelevinsky2006} & Theory, this work & Theory \cite{McGuyer2013} & This work & Exp. \cite{McGuyer2013} & Theory \cite{McGuyer2013} & Theory, this work \\
\hline
   $^{88}$Sr & $0_u^+$    & 1     & -0.435(37) & -0.418 & -0.427 &  16.5$^\circ$ & 17.46$^\circ$ & 0.666(14) & 0.636 & 0.659 \\
    & $0_u^+$    & 1     & -23.932(33) & -23.932 & -23.880 & 6.1$^\circ$ & 6.07$^\circ$ & 0.232(2) & 0.222 & 0.228 \\
    & $0_u^+$    & 1     & -222.161(35) & -222.162 & -222.167 &  4.2$^\circ$ & 4.04$^\circ$ & 0.161(2) & 0.148 & 0.147 \\
    & $1_u$    & 1     & -353.236(35) & -353.236 & -353.152 &  93.3$^\circ$ & 93.89$^\circ$ & 0.625(9) & 0.610 & 0.612 \\
    & $0_u^+$    & 1     & -1084.093(33) & -1084.092 & -1084.022 & 3.8$^\circ$ & 3.59$^\circ$ & 0.142(2) & 0.128 & 0.131 \\
    & $1_u$    & 1     & -2683.722(32) & -2683.723 & -2683.777 & 94.6$^\circ$ & 94.90$^\circ$ & 0.584(8) & 0.571 & 0.577\\
    & $0_u^+$    & 1     & -3463.280(33) & -3463.281 & -3463.346 & 5.1$^\circ$ & 4.88$^\circ$ & 0.193(3) & 0.174 & 0.173 \\
    & $1_u$    & 1     & -8200.163(39) & -8112.699 & -8200.219 & \{113.6$^\circ$, 175.9$^\circ$\}\footnote{McGuyer \emph{et al.} were uncertain which of the two -8~GHz states have $0_u^+$ and $1_u$ symmetries, and therefore gave mixing angles for both of the possible assignments.} & 112.58$^\circ$ & -0.149(2) & -0.592 & -0.024 \\
    & $0_u^+$    & 1     & -8429.650(42) & -8420.133 & -8427.888 & \{24.6$^\circ$, 84.9$^\circ$\}& 22.58$^\circ$ & 0.931(13) & 1.333 & 0.774 \\
    
      \noalign{\vskip 2mm}
    Isotope & Series & $J$     &Exp. \cite{McGuyer2013} &    & Theory, this work & Theory \cite{McGuyer2013} & This work & Exp. \cite{McGuyer2013} & Theory \cite{McGuyer2013} & Theory, this work \\ \hline
    $^{88}$Sr & $0_u^+$    & 3     & -0.63 & &   -0.644 & 18.9$^\circ$ & 18.92$^\circ$ & 0.270(2) & 0.271 & 0.271 \\
    & $0_u^+$    & 3     & -132  & & -134.217 & 11.6$^\circ$ & 12.22$^\circ$ & 0.173(2) & 0.160 & 0.157 \\
\noalign{\vskip 2mm}
    Isotope & Series  &   $J$    & Exp., this work & & Theory, this work & & This work & & & Theory, this work \\ \hline 
    $^{86}$Sr& $0_u^+$    & 1     & -1.633(10) & &  -1.534 & & 12.93$^\circ$ & & & 0.490\\
    & $0_u^+$    & 1     & -44.246(10) & &  -43.850 & & 6.36$^\circ$ & & & 0.214 \\
    & $1_u$    & 1     & -159.984(50) & & -159.993 & & 97.31$^\circ$ & & & 0.556 \\
    & $0_u^+$    & 1     & -348.742(10) & & -348.825 & & 6.67$^\circ$ & & & 0.206 \\

 \noalign{\vskip 2mm}
    Isotope & Series  &   $J$    & Exp. \cite{Stellmer2012} & & Theory, this work & & This work & & & Theory, this work\\ \hline 
    $^{84}$Sr & $0_u^+$ & 1     &    -0.320(10) & &  -0.296 & & 19.56$^\circ$ & & & 0.736\\
    & $0_u^+$    & 1     &   -23.010(10) & & -23.011 & & 6.48$^\circ$ & & & 0.242 \\
    & $0_u^+$    & 1     &  -228.380(10) & & -228.380 & & 3.47$^\circ$ & & & 0.116 \\
    & $0_u^+$    & 1     & -1288.290(10) & & -1144.492 & & 3.11$^\circ$ & & & 0.113 (0.101)\footnote{The number in parenthesis denotes the g-factor calculated using a potential fitted to this state alone.}\\
    \end{tabular}%
\end{ruledtabular}
  \label{tab:longrange}%
\end{table*}%

To perform photoassociation (PA), we prepare ultracold $^{86}$Sr atoms in an
optical dipole trap (ODT) via laser cooling and trapping techniques similar to
those used for other Sr isotopes \cite{Nagel2003,Yan2011a}. The ODT is formed by
the intersection of two mutually perpendicular beams focused to waists ($e^{-2}$
intensity radii) of 100 $\mu$m. Both beams are generated from a 1064-nm,
linearly-polarized, multilongitudinal-mode fiber laser. A period of forced
evaporation to a trap depth of 3.6\,$\mu$K yields $3 \times 10^5$ $^{86}$Sr
atoms at a temperature of 400\,nK and peak density of $10^{13}$\,cm$^{-3}$.

The PA beam is derived from a 689\,nm master-slave diode laser system that has a
linewidth of approximately 10 kHz. Short term stability is provided by locking
the laser frequency to a moderate finesse ($\mathcal{F}=2000$) optical cavity,
and long term stability is assured through saturated absorption spectroscopy of
the $^1S_0$-$^3P_1$ atomic transition in a vapor cell.  The PA beam is red
detuned with respect to the atomic transition using acousto-optic modulators
(AOMs) and transported to the atoms through a single-mode optical fiber. In the
interaction region, the beam is linearly polarized, with a waist of 700\,$\mu$m
and peak intensity up to 50\,mW/cm$^2$.  During the application of the PA beam,
to eliminate the ac-Stark shift due to ODT beams, the ODT is modulated with 50\%
duty cycle, a period of 462\,$\mu$s, and a peak trap depth of 7.2\,$\mu$K. The PA
beam is applied when the ODT is off. Total PA time is varied from 16 to 830\,ms,
depending upon the transition, to obtain a peak atom loss due to PA of
approximately 50\% with minimal change in sample temperature.  The number of
atoms and sample temperature are determined with time-of-flight absorption
imaging using the $^1$S$_0$-$^1$P$_1$ transition at 461\,nm.

New $^{86}$Sr photoassociation lines found in our experiment are listed in Table
\ref{tab:longrange}. The binding energies $\nu_0$ were obtained by fitting the
trap loss spectra with a realistic line shape function \cite{Ciurylo2004}. The
details of this procedure, the error budget and the respective optical lengths
are given in Appendix \ref{sect:lineshape}.

\section{Long range interactions \label{sec:longrange}}

\begin{figure}
\includegraphics[width=0.48\textwidth, clip]{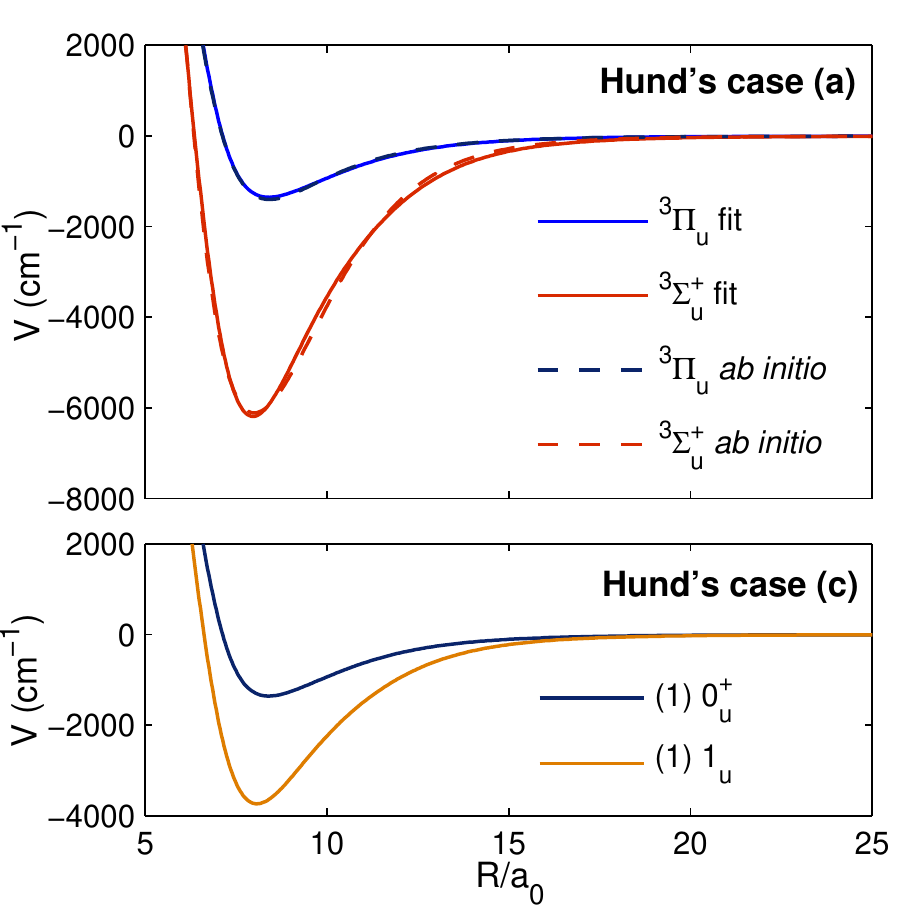}
\caption{(Color online) Potential curves correlating to the $^1S_0$+$^3P_1$
asymptote used in the calculation of the theoretical binding energies. The
dashed curves represent the original \emph{ab initio} Hund's case (a) potentials
of Skomorowski \emph{et al.} \cite{Skomorowski2012}. The solid lines are the
potential curves fitted to the experimental data from photoassociation
experiments. The top panel shows the potential curves in the Hund's case (a)
representation, while the bottom panel shows them in the Hund's case (c) basis
actually used in the calculation. \label{fig:longrange}}
\end{figure}

\begin{table}
	\caption{Potential parameters used in the calculation of theoretical
	$^1S_0$+$^3P_1$ bound state energies in Table \ref{tab:longrange}. The
	potential parameters given here are to be used with Eq.~\ref{eq:tt}. The only
	isotope-dependent parameter is $\alpha$, which is used to establish the correct
	short range quantum defect for each curve. \label{tab:parameters}}
	\begin{ruledtabular}
		\begin{tabular}{l r r}
		Parameter & $^3\Sigma_u^+$ & $^3\Pi_u$ \\
		\hline 
		$A_0$ & $  1.29406314 \times 10^2$ & $ 5.78723038 \times 10^6$\\
		$A_1$ & $ -7.90551852 \times 10^1$ & $-3.46113235 \times 10^6$\\
		$A_2$ & $  1.87863441 \times 10^1$ & $ 7.79019763 \times 10^5$\\
		$A_3$ & $ -1.96979418 \times 10^0$ & $-7.85317879 \times 10^4$\\
		$A_4$ & $  7.88636443 \times 10^{-2}$ & $ 3.01833743 \times 10^3$\\
		$\gamma$ & $7.61382806 \times 10^{-2}$ & $1.34967817 \times 10^{-3}$\\
		$\beta$ & $ 1.00 $ & $ 1.03238202 $\\
		$\alpha(^{88}\rm{Sr})$ & $ 0.045647282 $ & $1.99168225$\\	
		$\alpha(^{86}\rm{Sr})$ & $ 0.045690735 $ & $1.99188286$\\		
		$\alpha(^{84}\rm{Sr})$ & $ 0.045301189 $ & $1.99037413$\\
		$C_{12}$ & $-5.31841848 \times 10^9$ & $-1.06415514 \times 10^{10}$\\
		$C_{10}$ & $ 2.20495 \times 10^8$ & $5.24064 \times 10^7$\\
		$C_8$ & $2.3574797 \times 10^6$ & $3.4156471 \times 10^5$ \\
		$C_6$ & $4.3015063 \times 10^3$ & $3.8683912 \times 10^3$ \\
		$C_{3,0}$ & \multicolumn{2}{c}{$1.52356615 \times 10^{-2}$} \\
		\end{tabular}
	\end{ruledtabular}
\end{table}

In the original paper by Zelevinsky \emph{et al.} \cite{Zelevinsky2006}, the
energy levels obtained for $^{88}$Sr were modeled using a five channel model
\cite{Ciurylo2004} which included both the relevant molecular states from the
$^1$S$_0$+$^3$P$_1$ asymptote as well as ones from the $^1$S$_0$+$^1$P$_1$ and
$^1$S$_0$+$^3$P$_2$ asymptotes.  The energy levels for $^{88}$Sr
reproduced by this model are shown in Table \ref{tab:longrange} (labeled `Theory
\cite{Zelevinsky2006}').

In this paper we will use a two-channel model based on the two long range Hund's
case (c) potential curves, $0_u^+$ and $1_u$, that directly support the
vibrational states near the $^1S_0$+$^3P_1$ asymptote. The molecular Hamiltonian
can be partitioned in the following manner:
\begin{equation}
H = T+V_{\rm int}+V_{\rm rot}.
\end{equation}
Here, $T$ is the kinetic energy of the colliding atoms, $V_{\rm int}$ covers the
interaction potentials, while $V_{\rm rot}$ is the rotational energy of the
molecule. The kinetic energy term is diagonal regardless of chosen basis, with
the diagonal term $-({\hbar^2}/{2\mu})({d^2}/{dR^2})$. In our homonuclear case,
the reduced mass $\mu$ is equal to half the atomic mass of the chosen strontium
isotope.

In this paper we choose to work with potential curves based on state-of-the-art
\emph{ab initio} calculation of Skomorowski \emph{et al.} \cite{Skomorowski2012}
as opposed to using model potentials. Using the model in \cite{Mies1978} we can
write a two-channel Hund's case (c) atomic interaction hamiltonian in terms of
Hund's case (a) potential curves:
\begin{equation}
	V_{\rm int} = \left(
		\begin{array}{c c}
			\mathcal{V}(^3\Pi_u; R)-\frac{C_{3,0}}{R^{3}} & 0 \\
			0 & \frac{1}{2}\left(\mathcal{V}(^3\Pi_u; R)+\mathcal{V}(^3\Sigma^+_u; R)\right)-\frac{C_{3,1}}{R^{3}} 
		\end{array}
				\right) \,. \label{eq:CoriolisModel}
\end{equation}
Our Hund case (a) potentials $\mathcal{V}(^3\Pi_u, R)$ and
$\mathcal{V}(^3\Sigma_u^+, R)$ are based on the \emph{ab initio} potentials in
Skomorowski \emph{et al.}: $\mathcal{V}(^3\Pi_u, R) = \left(V({\rm c}~^3\Pi_u;
R)-V({\rm c}~^3\Pi_u; \infty)\right)$ and $\mathcal{V}(^3\Sigma_u^+, R) = V({\rm
a}~^3\Sigma^+_u; R)-V({\rm a}~^3\Sigma^+_u; \infty)$, where $V({\rm c}~^3\Pi_u;
R)$ and $V({\rm a}~^3\Sigma^+_u; R)$ are the respective potential curves. The
original \emph{ab initio} potential curves were given in the convenient form of
a short range part combined with a Tang-Toennies damped \cite{Tang1984} long
range part and enabled direct fitting of the potential parameters:
\begin{eqnarray}
	V(R)  =  e^{-\alpha R -\gamma R^2} 
			\left(A_0+A_1 R+A_2 R^2+A_3 R^3+A_4 R^4 \right) - \nonumber & &\\
			- C_{12} f_{12} R^{-12} - C_{10} f_{10} R^{-10} - C_8 f_8 R^{-8} -C_6 f_6 R^{-6} \, ,& & \label{eq:tt}
\end{eqnarray}
where $f_n (R)$ is a Tang-Toennies damping function of the {$n$-th} order
\cite{Tang1984}. During the fitting it was necessary to change the long range
$C_6$ and $C_8$ terms significantly. In order to retain the shape of the
potential curves we have refitted the remaining potential parameters to match
the shape of the original \emph{ab initio} potentials, as shown in
Figure~\ref{fig:longrange}. The potential parameters used in our calculations
are listed in Table~\ref{tab:parameters}. Finally, we have included the resonant
dipole interaction \cite{King1939} into the model. In Skomorowski \emph{et al.}
this was achieved by spin-orbit mixing between states correlating to the
$^1S_0$$+$$^1P_1$ and $^1S_0$$+$$^3P_1$ asymptotes. Since, however, we do not
expect any new physics emerging from the inclusion of the far $^1S_0$$+$$^1P_1$
asymptote, we decide to model this mixing by simply adding the dipole terms
artificially, following \cite{Mies1978}. In this case these terms are inversely
proportional to the lifetime of the $^3P_1$ atomic state in strontium:
\begin{equation}
	C_{3,0} = \frac{3}{2}\frac{\hbar}{\tau}\left(\frac{\lambda}{2\pi}\right)^3 \, ,
\end{equation}
 and 
\begin{equation}
	C_{3,1} = -C_{3,0}/2\,.\label{eq:dispersion}
\end{equation}
The resonant dipole interaction is
thus attractive in the $0_u^+$ curve while repulsive (and weaker) in
the $1_u$ state. 

 The remaining term in the molecular Hamiltonian is the
rotational energy $V_{\rm rot}$, which is diagonal in the Hund's case (e), but
not in the Hund's case (c) representation. This causes rotational (Coriolis)
mixing between the two molecular states:
\begin{equation}
	V_{\rm rot} = B(R) \left(\begin{array}{c c}
								J(J+1)+2 & -\sqrt{4J(J+1)} \\
								-\sqrt{4J(J+1)} & J(J+1) \\
							 \end{array}\right)
\end{equation}
with $B(R)={\hbar^2}/{2\mu R^2}$. In a similar study with ytterbium atoms
\cite{Borkowski2009}  the large dispersion between the two
potentials caused by the differences in the resonant dipole interaction (see
eq.(\ref{eq:dispersion})) made it possible to forego Coriolis mixing and fit
the available energy level data with a single channel potential. Here, however,
 the dipole interaction is much weaker, making it necessary to
include this mixing in order to properly recreate bound state energies very
close to the dissociation limit. Not including the Coriolis
mixing limits the accuracy of the model to about $1$~MHz for most energy levels.
The exception to this rule is the case of strongly mixed states, like the
$-8200$~MHz and $-8430$~MHz states in $^{88}$Sr, where this inaccuracy is
drastic,  as shown in Section \ref{sec:vibrational}.

Theoretical bound state energies can be obtained by solving the coupled channel
Schr\"odinger equation $H \Psi = E \Psi$, where $\Psi =
\left(\Psi(0_u^+; R),\,\Psi(1_u; R)\right)^T$ is the two-channel wavefunction. We
solve these equations numerically using the matrix DVR method
\cite{Tiesinga1998} with nonlinear coordinate scaling.

The long range parameters of the Hund's case (a) potentials $V({\rm
a}~^3\Sigma_u^+; R)$ and $V({\rm c}~^3\Pi_u; R)$ were fitted to the experimental
data using the nonlinear least-squares method. The long range resonant dipole
$C_3$, the two van der Waals terms $C_6({\rm a}~^3\Sigma_u^+)$ and $C_6({\rm
c}~^3\Pi_u)$ and the two respective $C_8$ terms were first used to match the
vibrational energies only for the $J=1$ $^{88}$Sr data. To a good approximation,
the $C_3$ and $C_6$ (and to a lesser extent $C_8$) coefficients determine the
vibrational splittings. Here, the short range $\alpha$ terms can be used to tune
the phases (or `quantum defects') of the short range parts of the radial
wavefunctions and effectively shift the whole vibrational series in place. This
parameter was chosen for phase adjustments in an attempt to preserve the shape
of the short range potential as much as possible. The resulting bound state
energies can be seen in Table \ref{tab:longrange}. The energy levels for $J=3$
were computed using the same set of parameters.

\begin{table}
\caption{Van der Waals coefficients $C_6$ for Sr$_2$ molecular states correlating to the $^1S_0$+$^3P_1$ asymptote given in Hund's case (a) and (c) bases.\label{tab:c6}}
\begin{ruledtabular}
\begin{tabular}{l l l l l}
Source & $C_6({}^3\Pi_u)$ & $C_6({}^3\Sigma_u^+)$ & $C_6(0_u^+)$ & $C_6(1_u)$ \\
\hline
Empirical \cite{Zelevinsky2006} & $3513 \pm 200$\footnotemark[1]  & $4035 \pm 600$\footnotemark[1]  & $3513 \pm 200$ & $3774 \pm 200$ \\
\emph{Ab initio} \cite{Mitroy2010} \footnotemark[3] & $3951$ & $4488$ & $3951$\footnotemark[2] & $4220$\footnotemark[2] \\
\emph{Ab initio} & & & & \\
(pure) \cite{Safronova2014} & 3821\footnotemark[1] & 4289\footnotemark[1] & 3821 & 4055 \\
\emph{Ab initio} & & & & \\
(recomm.) \cite{Safronova2014} & $3771 \pm 32$\footnotemark[1] & $4231 \pm 98$\footnotemark[1] & $3771 \pm 32$ & $4001 \pm 33$ \\
This work & $3868 \pm 50$ & $4302 \pm 150$ & $3868 \pm 50$ & $4085 \pm 50$ 
\footnotetext[1]{Calculated via $C_6({}^3\Pi_u)=C_6(0_u^+)$ and $C_6({}^3\Sigma_u^+)=2 C_6(1_u)-C_6(0_u^+)$ \cite{Mies1978} from Hund's case (c) values.}
\footnotetext[2]{Calculated via $C_6(0_u^+)=C_6(^3\Pi_u)$ and $C_6(1_u)=\left( C_6(^3\Sigma_u^+)-C_6(^3\Pi_u)\right)/2$ \cite{Mies1978} from Hund's case (a) values.}
\footnotetext[3]{Ref. \cite{Mitroy2010} did not give error bounds for the calculated values.}
\end{tabular}
\end{ruledtabular}
\end{table}

The vibrational level data for the remaining two isotopes, $^{86}$Sr and
$^{84}$Sr was modeled by adjusting only the two $\alpha$ coefficients on a per
isotope basis in order to fix the right short range wavefunction phase. The
differences between the $\alpha$ coefficients for different isotopes do not
exceed $1\%$. Therefore, the theoretical bound state energies listed in the
`Theory (this work)' of Table~\ref{tab:longrange} were calculated using slightly
different potential curves that shared the same set of the van der Waals $C_6$
and $C_8$ coefficients: $C_6(~^3\Sigma_u^+) \approx 4302~{\rm a.u.}$ and
$C_6(^3\Pi_u) \approx 3868~{\rm a.u.}$, which correspond to
$C_6(0_u^+)=3868~{\rm a.u.}$ and $C_6(1_u)=4085~{\rm a.u.}$ in the Hund's case
(c) representation. The $C_8$ coefficients are $C_8(^3\Sigma_u^+)\approx
2.36\times 10^6$ a.u. and $C_8(^3\Pi_u)\approx 3.42 \times 10^5$ a.u..  We
tentatively assign an error bound of $50~{\rm a.u.}$ to each of the Hund's case
(c) $C_6$ coefficients: a change of this size introduces a change in the
vibrational splittings that can not be compensated using the other long range
parameters. We choose to assign uncertainties to the Hund's case (c) parameters
because those directly affect the positions of the photoassociation resonances.
In the Hund's case (a) representation this results in uncertainties of 50~a.u.
and 150~a.u. for $C_6(^3\Pi_u)$ and $C_6(^3\Sigma_u^+)$ assuming our estimations
should be treated as maximum errors. Our fitted resonant dipole term $C_{3,0}
\approx 0.01524~{\rm a.u.}$ which corresponds to a natural linewidth of the
atomic $^3P_1$ state of $1/\tau=2\pi \times 7501.9$~Hz. Our lifetime of
$21.215~\mu$s agrees well with both the theoretical determination
\cite{Skomorowski2012} of $21.4~\mu$s and, not surprisingly, the empirical value
of $21.5~\mu${s} from the previous photoassociation experiment
\cite{Zelevinsky2006}.

Our $C_6$ coefficients can be compared to previous works (see Table
\ref{tab:c6}). The first photoassociation-based determination
\cite{Zelevinsky2006} gives $C_6(0_u^+)=3513~{\rm a.u.}$ and
$C_6(1_u)=3774$~a.u., respectively with an estimated error bound of $200$~a.u.
These values appear to be underestimated when compared to both our work and the
recent \emph{ab initio} determinations. The Hund's case (a) $C_6$ values
originally calculated by Mitroy and Zhang \cite{Mitroy2010} and used in the
model in Skomorowski et al. are $4488$~a.u. and $3951$~a.u. for the respective
$^3\Sigma_u^+$ and $^3\Pi_u$ states, which is about one and a half error bound
larger than our empirical determination. We note, however, that no error bound
was given for their calculation. A new \emph{ab initio} calculation by Safronova
\emph{et al.} \cite{Safronova2014} is also available. The first set of
coefficients, $C_6(0_u^+) = 3821$~a.u. and $C_6(1_u)=4055$~a.u. is of pure
\emph{ab initio} origin (labeled `\emph{Ab initio} (pure) \cite{Safronova2014}'
in Table \ref{tab:c6}) and fits our data perfectly. The second set of
coefficients (`\emph{Ab initio} (recomm.) \cite{Safronova2014}'), $C_6(0_u^+) =
3771(32)$~a.u. and $C_6(1_u)=4001(33)$~a.u., includes several empirical
corrections \cite{Safronova2013} and while the difference is larger, the data
still agree with ours to within mutual error bars. It should be noted that the
currently available Sr photoassociation data is only weakly sensitive to the
$C_8$ values. Therefore, $C_8(^3\Sigma_u^+)$ and $C_8(^3\Pi_u)$ should only be
considered as potential fitting parameters and not used for comparison with other
determinations. Not surprisingly, they do not agree with the newest available
\emph{ab initio} calculations \cite{Safronova2014}.

\section{Line positions \label{sec:vibrational}}

The agreement between our two-channel model and the experimentally determined
$^{88}$Sr bound state energies is excellent. The theoretical bound state
energies match the experimental line positions to within the error bars with the
exception of the two states at $-8430$~MHz ($0_u^+$) and $-8200$~MHz ($1_u$),
where the accuracy is limited to about $2$~MHz; see Table~\ref{tab:longrange}.
These two bound states are very strongly mixed by the Coriolis coupling,
partially due to the large wavefunction overlap. The two-channel wavefunctions
for these two energy levels are shown in Fig.~\ref{fig:coriolis}. Our
theoretical model predicts mixing angles, as defined in
Section~\ref{sec:zeeman}, of about $\theta \sim 23^\circ$, which is in good
agreement with the recent empirical determination based on Zeeman shifts of
photoassociation lines \cite{McGuyer2013}. Compared to the other energy levels,
quantitative description of these two states is very difficult, because the
Coriolis splitting between them is strongly dependent on the relative phases of
their respective wavefunctions which in turn are determined by the relatively
unknown short range parts of their supporting potential curves. We note that our
use of realistic \emph{ab initio} potentials improved the agreement
dramatically: the previous model \cite{Zelevinsky2006} was off by several tens
of MHz, while ours reduces that down to less than $2$~MHz. For completeness we
also show the theoretical counterparts of the two $J=3$ energy levels reported
recently in \cite{McGuyer2013}.

The same long range model (except for the adjustment of $C_8$
terms) applied to the case of $^{86}$Sr yields a slightly worse fit, with
inaccuracies reaching up to $0.5$~MHz, which can be attributed to the still
imperfect description of Coriolis mixing (in the case of the top bound state)
or to the impact of short range curve crossings on the vibrational splittings as
discussed in detail in Section~\ref{sec:spacings}.

The case of $^{84}$Sr atoms has been experimentally investigated recently by
Stellmer \emph{et al.} \cite{Stellmer2012}. In this analysis we, however, leave
out the $1_u$ state at $-351$~MHz state reported in \cite{Stellmer2012}, but not
in the energy level table in a subsequent review paper \cite{StellmerPrivate,
Stellmer2014}.  Only bound states of $0_u^+$ symmetry are therefore available
for the case of $^{84}$Sr atoms.

The experimental bound state energies for the $^{84}$Sr $0_u^+$ symmetry
obtained by Stellmer \emph{et al.} \cite{Stellmer2012} have vibrational spacings
that can not be fully reproduced by our long range model. Given the excellent
agreement between theory and experiment in the case of the two other isotopes we
can safely assume that at least the two van der Waals parameters $C_6(^3\Pi_u)$
and $C_6(^3\Sigma_u^+)$ and the resonant dipole term $C_3$ of the two potentials
are correct. Similarly to the case of $^{86}$Sr we have adjusted the two
$\alpha$ terms to fit the model to the $^{84}$Sr data. Due to the lack of
available $1_u$ bound state data, the $\alpha(^3\Sigma_u^+)$ parameter was
fitted to provide the best fit of $0_u^+$ energy levels. If we fit the model to
all of the $0_u^+$ bound states except for the $-1288$~MHz level, we obtain a
fit to better than $25$~kHz for all but the $-1288$~MHz state. This is shown in
Table \ref{tab:longrange}. The latter state appears not to follow the $J=1$
$0_u^+$ series, as the closest energy level predicted by our long range model is
located at about $-1144$~MHz, that is, almost $150$~MHz away. On the other hand,
if we instead fit our $0_u^+$ series to the $-1288$~MHz state we obtain a model
that is in drastic disagreement with the remaining experimental bound state
positions giving line positions of -0.57, -29.3 and -269.3 MHz.

Our long range model's inability to describe the $^{84}$Sr vibrational spectrum
suggests that the $-1288$~MHz state could be either perturbed by an adjacent
state in a different potential curve or it could have been mislabeled as a $J=1$
$0_u^+$ state. In fact, given that the experiment \cite{Stellmer2012} was
performed in a Bose-Einstein condensate it is plausible that this
photoassociation line is supported by one of the subradiant \emph{gerade} states
($0_g$ or $1_g$), much like those observed in ytterbium \cite{Takasu2012}.
In this case it would be entirely unperturbed by the $0_u^+$ and $1_u$ series due
to symmetry. This theory could be confirmed by actually finding an unperturbed
$0_u^+$ state near $-1144$~MHz as predicted by the long range model. 
It is important to note that \emph{not} finding such state does not necessarily
disprove its existence -- the intensity of a PA line can be greatly diminished
if the ground state wavefunction has a node at the Condon point for such
transition \cite{Julienne1996, Takasu2012}.

We have verified that the apparent shift of the $-1288$~MHz state is not caused
by a simple rotational state labelling error. In such scenario this energy level
could indeed have $J=1$, but the remaining three $0_u^+$ states could have $J=3$
and therefore lay closer to the dissociation limit. However, if this were the
case the theoretical bound state energies would be $-159$~MHz and $-1.5$ MHz for
$J=3$ with the most weakly bound state disappearing altogether.

We have tested a possibility that the $-1288$ MHz state belongs to the $1_u$
symmetry, which is plausible as so far no $1_u$ resonances were found in
$^{84}$Sr. However, if we fit the $\alpha(^3\Sigma_u^+)$ parameter so that our
model reproduces a $1_u$ state at $-1288$ MHz, another bound state of this
symmetry emerges at $-2.90$ MHz. Its presence causes a Coriolis shift of the top
weakly bound $0_u^+$ state  to $-0.21$ MHz, that is, over 10 error bounds away
from the experimental value of $-0.320(10)$ MHz. Therefore we view such
possibility as unlikely.

Finally, the $0_u^+$ vibrational spacings could be influenced by the strong
short-range spin-orbit mixing with other electronic states. For example, the
original Mies \emph{et al.} model \cite{Mies1978} contains non-diagonal
spin-orbit terms between $1_u$ curves correlating to the $^1S_0$$+$$^3P_1$ and
$^1S_0$$+$$^3P_2$ asymptotes. Such strong mixing could create very wide
resonances spanning several bound states near the $^1S_0$$+$$^3P_1$ dissociation
limit. Such a case could occur in all PA spectroscopy experiments involving
divalent atoms like ytterbium or calcium, but to the best of our knowledge, no
empirical evidence has so far been found. The impact of this mixing can be
somewhat diminished by the fact that potential curves of different $j$-states of
the same $^3P$ asymptote are largely parallel and do not cross. In strontium,
however, the situation is further complicated by an additional $^1S_0$$+$$^1D_2$
$0_u^+$ curve crossing the $0_u^+$ state probed in photoassociation experiment.
We will explore its consequences in Section~\ref{sec:massscaling}.

\section{Coriolis mixing and Zeeman splittings \label{sec:zeeman}}

\begin{figure}
\includegraphics[width=0.48\textwidth, clip]{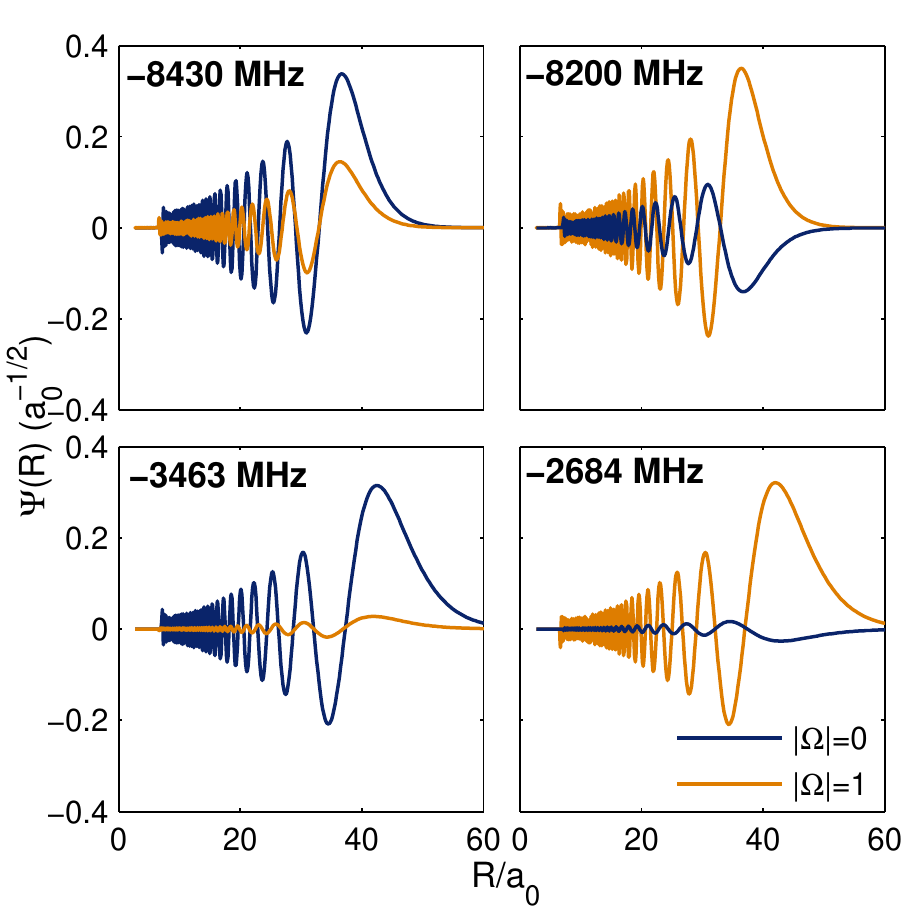}
\caption{ (Color online) Coriolis mixing between the $0_u^+$ and $1_u$
series. The upper two plots show the two-channel wavefunctions of the strongly
mixed $^{88}$Sr $J=1$ levels at $-8430$~MHz and $-8200$~MHz ($\theta=24^\circ$
and $\theta=114^\circ$, respectively). In the case of the $-8430$~MHz level, the
$0_u^+$ component (dark blue lines) constitutes the majority of the two-channel
wavefunction, while the $-8200$~MHz state is predominantly of the $1_u$ symmetry
(orange lines). For comparison, in the lower part of the figure we show two
relatively pure ($\theta = 5^\circ$ and $\theta = 95^\circ$) $0_u^+$ and $1_u$
states at $-3463$~MHz and $-2684$~MHz, respectively. \label{fig:coriolis}}
\end{figure}

The Coriolis terms in the rotational Hamiltonian cause nonadiabatic mixing
between the $0_u^+$ and $1_u$ components of the molecular wavefunction. This
effect can be quantified by introducing a mixing angle $\theta$ defined by
writing the total molecular wavefunction as:
\begin{equation}
	\Psi(R) = \left(\cos \theta \, \tilde \Psi(0_u^+; R), \sin \theta \, \tilde \Psi(1_u; R) \right)^T \,. \label{eq:angle}
\end{equation}
Here we define the two reference functions $\tilde \Psi(0_u^+; R)$ and $\tilde
\Psi(1_u; R)$, which are the two wavefunction components normalized separately
via:
\begin{equation}
	\tilde \Psi(0_u^+; R) = \Psi(0_u^+; R) \left(\int_0^\infty \Psi(0_u^+; R)^2 dR \right)^{-1/2} \, ,
\end{equation}
and 
\begin{eqnarray}
	\tilde \Psi(1_u; R) & = & \Psi(1_u; R) \left(\int_0^\infty \Psi(1_u; R)^2 dR \right)^{-1/2} \times \nonumber \\
	& & \times \, {\rm sgn}  \left(\int_0^\infty \Psi(0_u^+; R) \Psi(1_u; R) dR \right) \, .
\end{eqnarray}
The signum function above ensures that our phase convention is compatible with
the one in Ref. \cite{McGuyer2013}. It is straightforward to verify that with
these definitions Eq.~\ref{eq:angle} yields a correctly normalized two-channel
wavefunction. Finally, we note that the transformation $\theta \rightarrow
\theta + 180^\circ$ only changes the sign of the total wavefunction without
altering its internal phase relationship. We therefore decide on having $0^\circ
\leq \theta < 180^\circ$, again in accordance with \cite{McGuyer2013}. In this
convention, a pure $0_u^+$ state has $\theta = 0^\circ$ or $\theta \to
180^\circ$, while $\theta = 90^\circ$ for a pure $1_u$ state.

Table \ref{tab:longrange} lists our Coriolis mixing angles $\theta$ for each of
the considered energy levels. For completeness we also include mixing angles
from Ref.~\cite{McGuyer2013} which are in good agreement with ours. For two of
the most deeply bound states in $^{88}$Sr at $-8200$ MHz and $-8430$~MHz,
McGuyer~\emph{et al.} give two different mixing angles as their theoretical
model was not accurate enough to ascertain which of these two energy levels
belong to the $0_u^+$ and $1_u$ series. In our model, however, the $-8200$~MHz
line clearly belongs to the $1_u$ series, while the $-8430$~MHz has $0_u^+$
symmetry, confirming the original assignment of Zelevinsky~\emph{et
al.}~\cite{Zelevinsky2006}.

Coriolis mixing is a significant factor in the bound state energies close to the
$^1S_0$+$^3P_1$ limit in strontium. Not surprisingly, the energy shift by this
mixing is dependent on the mixing angle $\theta$. This is especially important
for the very long range top bound states: for example, without Coriolis mixing
the theoretical energy for the $-0.435$~MHz level in $^{88}$Sr is $-0.160$~MHz.
An even more extreme case is the top bound state in $^{84}$Sr where the
theoretical binding energy would be $-37$~kHz as opposed to the experimental
value of $-0.32$~MHz. Moreover, Coriolis mixing is especially strong when two
energy levels of $0_u^+$ and $1_u$ coincide, as is the case of the two energy
levels in $^{88}$Sr at $-8200$ MHz and $-8430$~MHz. The mixing angle
$\theta=23^\circ$ for the $-8430$~MHz state, indicating particularly strong
mixing, which again significantly influences the binding energy. In fact,
without Coriolis mixing the respective theoretical energies are $-8234$~MHz and
$-8394$~MHz, missing the experimental line positions by over $30$~MHz. These
two states are shown in the upper part of Fig.~\ref{fig:coriolis}. The remaining
bound states have mixing angles of about $5^\circ$ for $0_u^+$ states and $\approx
95^\circ$ for $1_u$ symmetry and therefore are relatively pure, as shown for the
$-3463$~MHz ($0_u^+$) and $-2684$~MHz ($1_u$) states in the lower part of
Fig.~\ref{fig:coriolis}.

Our theoretical model provides improved description of the nonadiabatic Coriolis
mixing between the $0_u^+$ and $1_u$ states. Recently, McGuyer {\emph{et al.}}
\cite{McGuyer2013} gave experimentally determined linear Zeeman coefficients $g$
for the $^{88}$Sr photoassociation lines, as well as their theoretical
counterparts. The Zeeman coefficients
\begin{equation}
	g = g_{at} \left( \frac{\sin^2 \theta}{J(J+1)} + \frac{\sin  2\theta}{\sqrt{(J(J+1))}} P_{01} \right) \label{eq:gfactor}
\end{equation}
are highly sensitive both to the mixing angle $\theta$ and the overlap $P_{01}$
of the components $\tilde \Psi(0_u^+; R)$ and $\tilde \Psi(1_u; R)$:
\begin{equation}
  P_{01} = \int_0^\infty \tilde \Psi(0_u^+; R) \tilde \Psi(1_u; R) dR \, .
\end{equation}
The atomic $g$-factor $g_{at} = 1.5$ for the $^3P_1$ electronic state.

A comparison of experimental and theoretical Zeeman $g$-factors from
Ref.~\cite{McGuyer2013} with our values calculated with Eq.~\ref{eq:gfactor} is
given in Table~\ref{tab:longrange}. For relatively pure $0_u^+$ and $1_u$ energy
levels our $g$-factor agree very well with the theoretical values in
\cite{McGuyer2013}. In the case, however, of the top bound state at $-0.4$ MHz
our theoretical value is slightly closer to the one obtained in experiment. The
most striking improvement is seen in the case of the two strongly Coriolis-mixed
bound states at $-8200$ MHz and $-8430$~MHz. As noted previously, these two
energy levels are notoriously difficult to describe theoretically and the model
in \cite{McGuyer2013} fails to reproduce their experimental $g$-factors. Our
model reduces the discrepancy between theory and experiment by a factor of three
and while our model still does not fit to within experimental accuracy, it at
least gives qualitative agreement. Since our mixing angles $\theta$ are in very
good agreement, we attribute this improvement to the better description of the
wavefunction overlap $P_{01}$, which was a necessary condition to correctly
reproduce the impact of Coriolis mixing on the positions of these two energy
levels. This further corroborates the validity of our long range van der Waals $C_6$, $C_8$ and resonant dipole $C_3$
coefficients. For completeness we also provided our Zeeman $g$-factors for the
remaining PA lines, but no experimental data is currently available to compare.

\section{Mass scaling perturbed \label{sec:massscaling}}

\subsection{Single channel mass scaling \label{sec:massscaling_singlechannel}}

A mass scaled model is an interaction model that is capable of
reproducing the energy levels for all isotopomers of a given molecule by only
changing the reduced mass. By its nature photoassociation spectroscopy is
relatively insensitive to the details of the short range atomic interaction as
it is predominantly used to measure the energies of bound states very close to
the dissociation limit. This can be understood using the following simple
reasoning. The energy splittings between the vibrational states (on the order of
$1$~GHz) are very small compared to the depth of the potential curve (tens of
THz). Consequently, bound states close to the dissociation limit share the same
short range wavefunction (to within an amplitude constant) and only differ in
the long range, where even such small energy differences will influence the
location of the outer classical turning point and the long range wavefunction.
Therefore, to a certain approximation, the vibrational splittings would be
determined by the long range potential parameters, like $C_6$ and $C_3$ in our
case. On the other hand, the starting point of the vibrational
state ladder would be determined by the vibrational wavefunction's phase $\phi$,
as required by the Bohr-Sommerfeld quantization condition. Such
a philosophy is the basis for the famous analytic expression for the $s$-wave
scattering length of an $R^{-n}$ potential \cite{Gribakin1993}, as well as the
semiclassical Leroy-Bernstein formulas \cite{LeRoy1970}.

As a consequence of the above reasoning, obtaining mass scaling in a single
channel model would only require a potential curve with the correct long range
part and the right behavior of the total phase $\phi$ as a function of the
reduced mass. Such an approach has been discussed in detail in
\cite{Borkowski2013}. In the case of a single potential curve the total phase
can be well approximated using the WKB integral:
\begin{equation}
	\phi = \frac{1}{\hbar} \int_{R_{\rm in}}^{\infty} \sqrt{-2\mu V(R)}
	dR \,, \label{eq:wkbphase}
\end{equation}
where $R_{\rm in}$ is the location of the inner classical turning point and
$V$ is the interaction potential curve. It is evident that in
this approach the total phase $\phi$ is proportional to $\sqrt{\mu}$ as long as the
potential $V$ is mass-independent. In this paper we will assume that the potential $V$ is the same for all isotopes.

Following the Bohr-Sommerfeld quantization condition the quantity $n
= \phi/\pi$ is closely related to the number of bound states supported by the
interaction potential. On the other hand, the positions of the vibrational
states only depend on the fractional part $\Delta n = n - \lfloor
n \rfloor$. Slowly increasing $n$, by e.g. adjusting the depth of the potential
curve, causes the bound state energies to shift deeper into the potential well. At some
point this will cause a new bound state at the dissociation limit to emerge.
Increasing $n$ by 1 (or, equivalently, the total phase $\phi$ by
$\pi$) would amount to adding exactly one vibrational state to the potential,
with the weakly bound states having similar vibrational energies as previously.
If we assume that the interaction potential $V$ is mass-independent, then the
quantity $n$ is proportional to $\sqrt{\mu}$. In this picture, mass scaling
would amount to finding a potential curve which offers correct values of $\Delta
n$ for each isotope. This can be achieved by simply fixing the correct number of
bound states supported by the interaction potential. Such a strategy has been
successfully implemented in a number of PA investigations \cite{Kitagawa2008,
Borkowski2009, Borkowski2013}.

In our case the long range interactions near the $^1S_0$$+$$^3P_1$ asymptote are
described by two Hund's case (c) potentials, $0_u^+$ and $1_u$ coupled by Coriolis
mixing. We note, however, that this mixing is relatively weak and only provides
a small correction to most bound state energies. In fact, foregoing the Coriolis
mixing altogether does not change the number of bound states supported by the
two potentials together. Therefore, mass scaling of both series can to some
extent be treated separately. We will first consider the mass scaling of energy levels of $0_u^+$ symmetry, for which most of the experimental data was collected. The final step will be the addition of the $1_u$ states to the model.

\begin{figure}
	\includegraphics[width=0.48\textwidth,clip]{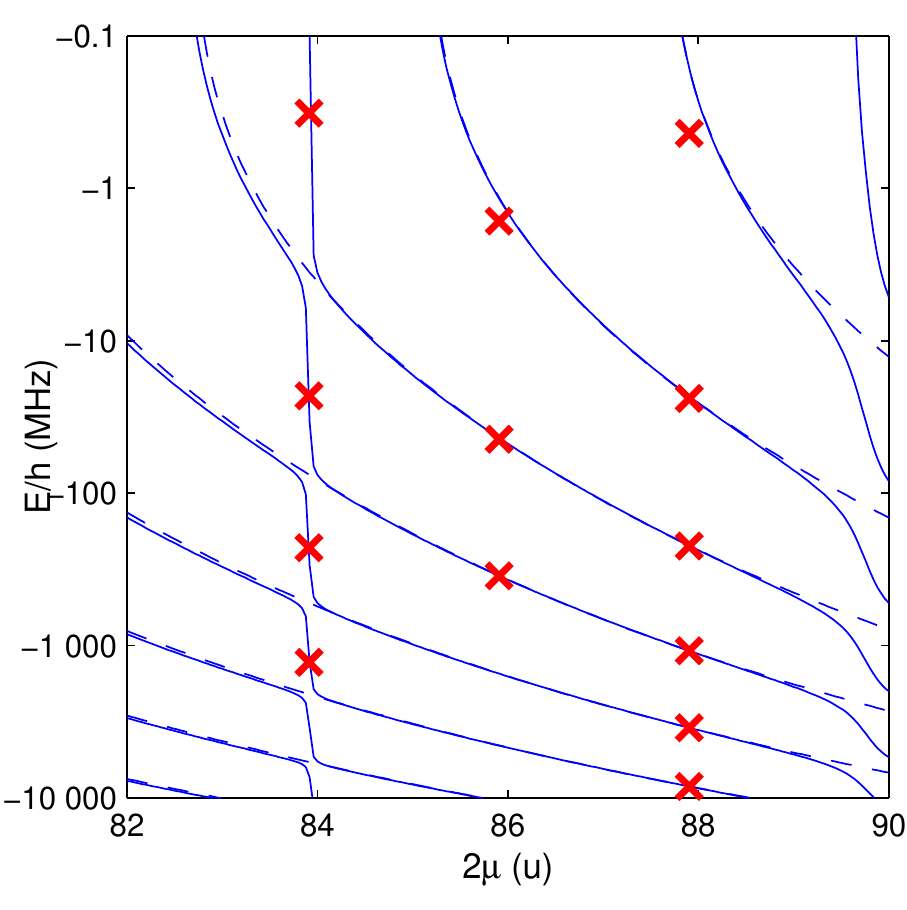}
	\caption{(Color online) Theoretical $0_u^+$ energy levels as a function of the
	reduced mass $\mu$. The blue dashed lines are calculated using the $^3\Pi_u$
	potential curve alone which is only capable of mass scaling between two
	isotopes at a time --- $^{86}$Sr and $^{88}$Sr in this case. The solid lines
	show the results for a model that also includes the short range crossing with a
	$^1S_0$+$^1D_2$ $^1\Sigma_u^+$ potential, as shown in Fig.~\ref{fig:crossing}.
	The latter model is capable of reproducing the experimental bound state
	energies (marked as red crosses) for all isotopes. In this model, a short range
	$^1\Sigma_u^+$ perturbing state crosses the $^1S_0$$+$$^3P_1$ dissociation
	limit at $2\mu \approx 84$~u causing a sudden departure from the usual mass
	scaling behavior.
	\label{fig:massscaling}}
\end{figure}

Figure \ref{fig:massscaling} shows the $0_u^+$ bound state energies as a function
of the reduced mass. By changing the $\alpha(^3\Pi_u)$ potential parameter we
have found an appropriate $0_u^+$ potential depth (and consequently $\phi$) that
correctly translated the wavefunction phase between the isotopes $^{86}$Sr and
$^{88}$Sr. The results are shown in Fig. \ref{fig:massscaling} as dashed lines.
We have found that while our model fits the experimental bound state energies
for $^{86}$Sr and $^{88}$Sr to within $1$~MHz (with the
exception of the strongly Coriolis mixed $-8430$ MHz state), it is in shocking
disagreement with the experimental data for $^{84}$Sr.

This clear failure of mass scaling be easily explained \emph{a
posteriori} using the simple picture described earlier. The fractional part
$\Delta n$ of the total WKB phase $\phi/\pi$ of the $0_u^+$ state (computed by
applying Eq. \ref{eq:wkbphase} to $\mathcal V(^3\Pi_u; R)$) for the best-fit
models from Section~\ref{sec:longrange} is $\Delta n = 0.3928$, $0.6003$ and
$0.3245$ for $^{88}$Sr, $^{86}$Sr, and $^{84}$Sr, respectively. In the case of a
single potential curve, $n$ scales linearly with $\sqrt{\mu}$. The model shown
in Fig. \ref{fig:massscaling} shows bound state energies calculated using a
potential that has $n=69.3928$ and mass scales, correctly, to $n=68.6003$ for
$^{86}$Sr, which fits the observed line positions. Note that this model supports
one bound state fewer for the latter isotope. For the $^{84}$Sr isotope,
however, mass scaling gives $n=67.7988$, but this fails to predict the observed
line positions.

Similar mass scaling can be obtained for other numbers of bound states, but in
all cases only for two isotopes at a time. For example, in the physically
reasonable range of $50$ to $150$ supported bound states, mass scaling between
$^{88}$Sr and $^{86}$Sr is obtained only for $69$ vibrational states, while for
$^{88}$Sr and $^{84}$Sr the correct phases can be obtained for $89$ or $133$
states. Finally, for the $^{86}$Sr-$^{84}$Sr pair a potential curve with $111$
bound states would support mass scaling. Clearly none of these numbers overlap.
In fact, within this simple approximation no single potential curve supporting
less than $300$ vibrational states could support mass scaling between all three
isotopes. For comparison, an unmodified $^3\Pi_u$ curve from
\cite{Skomorowski2012}, as used in our model in Section \ref{sec:longrange}, 
supports a total of $72$ bound states for $J=1$ in $^{88}$Sr.

\subsection{The avoided crossing}

\begin{figure}
	\includegraphics[width=0.48\textwidth]{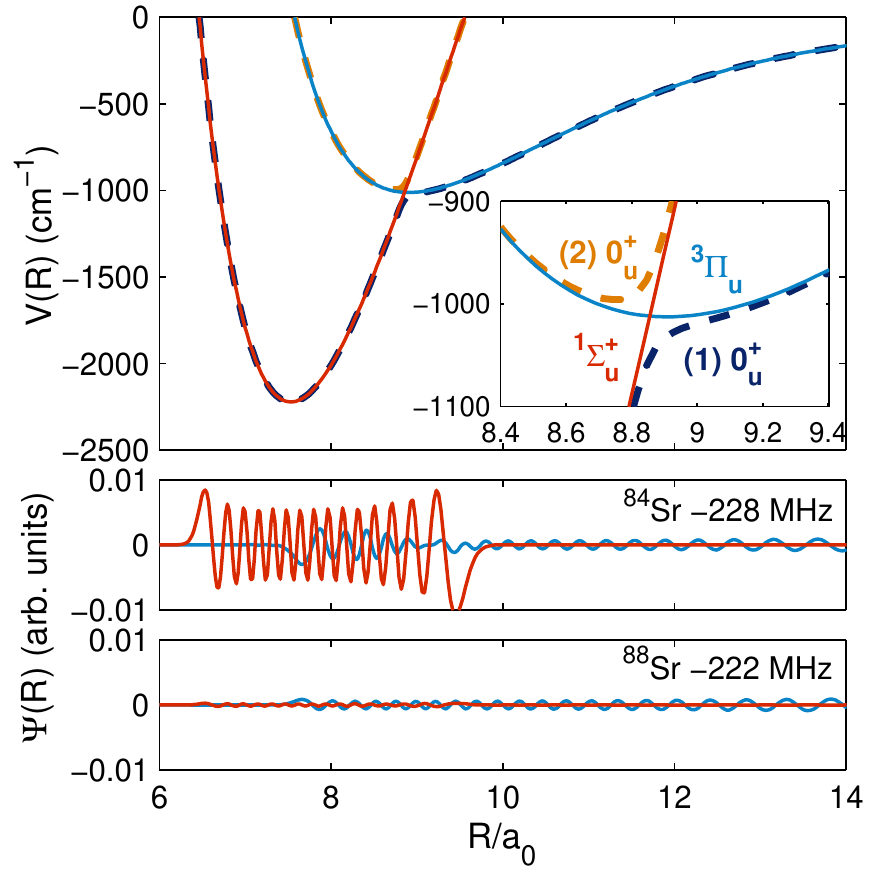}
	\caption{(Color online) Avoided crossing between the $^1$S$_0$+$^3$P$_1$
	$^3\Pi_u$ and $^1$S$_0$+$^1D_2$ $^1\Sigma_u^+$ potential curves. The strong
	spin-orbit coupling between these two potential curves causes anomalies in the
	mass-scaling behavior of the photoassociation spectra near strontium
	intercombination line. In fact, a model that only includes the potential curve
	directly supporting the $0_u^+$ bound states close to the $^1$S$_0$+$^3$P$_1$
	dissociation limit is incapable of properly describing mass scaling between the
	three bosonic isotopes in strontium.
\label{fig:crossing}}
\end{figure}

\begin{table}
  \centering
  \caption{A comparison of experimental $0_u^+$ bound state energies
  with predictions of our mass-scaled model that includes the strong spin-orbit
  coupling between the $^1S_0$+$^3P_1$ $^3\Pi_u$ curve which directly supports
  the $0_u^+$ series observed in PA experiments and the $^1S_0$+$^1D_2$
  $^1\Sigma_u^+$ curve. See text.}
\begin{ruledtabular}
    \begin{tabular}{c c c c c c c}
          & \multicolumn{2}{c}{$^{88}$Sr} & \multicolumn{2}{c}{$^{86}$Sr} & \multicolumn{2}{c}{$^{84}$Sr}  \\
    $\nu$     & Exp.  & Theory & Exp.  & Theory & Exp.  & Theory \\
    \hline
    -1    & -0.435 & -0.164 & -1.633 & -1.169 & -0.320 & -0.071 \\
    -2    & -23.932 & -23.133 & -44.246 & -43.467 & -23.010 & -23.208 \\
    -3    & -222.161 & -221.137 & -348.742 & -349.115 & -228.380 & -228.173 \\
    -4    & -1084.093 & -1083.353 &       &       & -1288.290 & -1062.636 \\
    -5    & -3463.280 & -3463.857 &       &       &       &  \\
    -6    & -8429.650 & -8410.027 &       &       &       &  \\
    \end{tabular}%
\end{ruledtabular}
  \label{tab:massscaling}%
\end{table}%

It has recently been established, both experimentally \cite{Stein2010} and
theoretically \cite{Skomorowski2012}, that the $^3\Pi_u$ potential which
supports the Hund's case (c) $0_u^+$ potential correlating to the
$^1$S$_0$+$^3$P$_1$ asymptote has a strong short range avoided crossing with a
$^1\Sigma_u^+$ curve which corresponds to the much higher $^1$S$_0$+$^1$D$_2$
asymptote.  The resulting $^1$S$_0$+$^3$P$_1$ $0_u^+$ curve, as shown in
Fig.~\ref{fig:crossing}, has several intriguing characteristics: its depth is
defined almost solely by the depth of the $^1\Sigma_u^+$ potential and it
enables nuclear motion at shorter distances than the $^3\Pi_u$ curve alone.
Skomorowski \emph{et al.}\cite{Skomorowski2012} have analyzed the composition of
bound states in the $0_u^+$ state. The deepest bound states in the potential
well are comprised primarily of the $^1\Sigma_u^+$ state, as expected given
their energies are below the minimum of the $^3\Pi_u$ curve. As we move up the
bound state ladder, the bound states become seemingly erratic mixtures of the
$^1\Sigma_u^+$ and $^3\Pi_u$ states. Finally, the bound states closest to the
dissociation limit (and so far investigated by photoassociation spectroscopy)
are nearly pure $^3\Pi_u$ states, which explains why our long
range model alone was enough to quantitatively describe most of the observed
vibrational spacings.

To test the influence of the $^{1}\Sigma_u^+$ state on the $^{3}\Pi_u$ bound
states and mass scaling we use a two channel model that includes both
potentials:

	\begin{equation}
		V_{\rm int} = \left( \begin{array}{c c}
						\mathcal{V}(^3\Pi_u;R)-C_{3,0}R^{-3}&
						 \sigma \xi_1(R) \\
						\sigma \xi_1(R) &
						 \mathcal{V}(^1\Sigma_u^+;R) + \Delta E\\
					\end{array}
			\right) \, ,
	\end{equation}
where, similar to Section \ref{sec:longrange}, $\mathcal{V}(^1\Sigma_u^+;R) =
{V}({\rm A} ^1\Sigma_u^+;R) - {V}({\rm A} ^1\Sigma_u^+;\infty)$. This model
obviously lacks the $1_u$ curve, and consequently, it does not include Coriolis
mixing, limiting its accuracy to about $1$~MHz for bound states with small
Coriolis mixing angles. The diagonal rotational terms are $B(R)(J(J+1)+2)$ and
$B(R)J(J+1)$, respectively. The potential $V(^3\Pi_u;R)$ has the same parameters
as the one from Section~\ref{sec:longrange} except for $\alpha(^3\Pi_u)$, which
we use to adjust the quantum defect. The ${V}({\rm A} ^1\Sigma_u^+;R)$ potential
and spin-orbit coupling function $\xi_1(R)$ is the same as in
\cite{Skomorowski2012}. Again, we model the influence of the $^1$S$_0$+$^1$P$_1$
$^1\Sigma_u^+$ curve by manually adding the resonant dipole $-C_{3,0}R^{-3}$
term. The two additional fitted parameters are $\sigma \approx 0.42$, which
enabled the scaling of the spin-orbit mixing between the two potential curves,
and $\Delta E \approx 6250~{\rm cm}^{-1}$ which controls the splitting between
the $^3$P$_1$ and $^1$D$_2$ atomic states. We use the latter parameter to fix
the position of the perturbing short range $^1\Sigma_u^+$ bound state with
respect to the $^1S_0$$+$$^3P_1$ asymptote while retaining the original shape of
the $^1\Sigma_u^+$ potential. As was the case previously (Section
\ref{sec:longrange}), the theoretical energy levels are obtained by solving the
coupled channel Schr\"odinger equation for the above potential matrix.

The solid lines in Figure~\ref{fig:massscaling} show the theoretical bound state
energies calculated using the above two-channel model fitted to the experimental
$0_u^+$ data for all isotopes using only three parameters: $\alpha(^3\Pi_u)$,
which as previously we used to adjust the short range wavefunction
phase, the mixing parameter $\sigma$ that scales the \emph{ab initio}
\cite{Skomorowski2012} spin-orbit coupling function and the relative positions
of the two potential curves $\Delta E$. The long range parameters,
$C_6(^3\Pi_u)$ and $C_{3,0}$ are shared with the previously discussed long
range model from Section~\ref{sec:longrange}. Apart from the $^{84}$Sr
$-1288$~MHz state discussed in detail in Section~\ref{sec:vibrational}, and the
strongly Coriolis mixed $-8430$~MHz state in $^{88}$Sr, the energy levels are
reproduced to within $1$~MHz. We note that the fitting itself was a very
difficult process due to the seemingly erratic behavior of the model.
This is caused by the mixing between the two diabatic potential
curves $^3\Pi_u$ and $^1\Sigma_u^+$ being strongly dependent on the relative
phases of the two components $\Psi({}^3\Pi_u)$ and $\Psi({}^1\Sigma_u^+)$ of
the coupled channel wavefunction. This is shown in the bottom part of
Fig.~\ref{fig:crossing}. In our fit the $-222$~MHz state of $^{88}$Sr is a
nearly pure $^3\Pi_u$ state, while the $-228$~MHz level in $^{84}$Sr is
strongly influenced by a short range $^1\Sigma_u^+$ component. Such a dramatic
change of $^1\Sigma_u^+$ component amplitude is likely the result of a bound
state of this symmetry coinciding with the $^1S_0$+$^3P_1$ asymptote in
$^{84}$Sr.

The mass scaling behavior of this model (solid lines in
Fig.~\ref{fig:massscaling}) is similar to the single channel case (dashed lines)
for the range of reduced masses of about $2\mu \approx 85\ldots89$~u. However,
when a short range $^1\Sigma_u^+$ bound state crosses the $^1S_0$$+$$^3P_1$
dissociation limit, there is a sudden resonant departure as seen near $2\mu
\approx 84$~u and $2\mu \approx 90$ u. Since the spin-orbit mixing is
relatively strong, the width of this resonant structure is large and
encompasses all bound states close to the dissociation limit together, resulting
in an apparent change in the quantum defect. This explains why the vibrational
spacings alone can be described by a single channel  model even
when the mixing is large. A comparison of the experimental $0_u^+$ energy levels
with ones calculated from our mass scaled model is shown in Table
\ref{tab:massscaling}. The accuracy of the model is about $1$~MHz for the
majority of the experimental data and is mostly limited by the lack of Coriolis
mixing. 

\subsection{Impact on vibrational spacings \label{sec:spacings}}

\begin{table}

\caption{A comparison of theoretical energy levels of $0_u^+$ symmetry calculated
with (`Perturbed') and without (`Adiabatic') the spin-orbit mixing
between the $^1\Sigma_u^+$ and $^3\Pi_u$ curves. The quantum defect has been 
adjusted in the adiabatic case so that the energy of the $\nu=-3$ bound state 
strictly matches the one predicted by the two-channel model. This ensures that
any differences between the energy level predictions are caused by the way the
models are constructed rather than quantum defects. In our model the $^{84}$Sr 
bound states close to
the dissociation limit are significantly perturbed by this mixing, which changes
their vibrational splittings significantly. On the other hand the $^{88}$Sr
bound states remain relatively unchanged. This behavior can be used in the
future to ascertain the location of the perturbing states. All values are given
in MHz.\label{tab:testing}}

\begin{ruledtabular}
	\begin{tabular}{r r r r r}
		& \multicolumn{2}{c}{$^{84}$Sr} & \multicolumn{2}{c}{$^{88}$Sr} \\
		$\nu$ & Perturbed & Adiabatic & Perturbed & Adiabatic \\
		\hline	
		-3 & -228.4 & -228.4 & -222.2 & -222.2 \\
		-4 & -1063.3 & -1147.6 & -1087.0 & -1086.8 \\
		-5 & -3087.8 & -3704.1 & -3472.5 & -3471.2 \\
		-6 & -6983.4 & -9024.2 & -8426.4 & -8419.9 \\
		-7 & -13987.8 & -18316.1 & -17080.6 & -17056.0 \\		
	\end{tabular}
\end{ruledtabular}
\end{table}

The available photoassociation data for $J=1$ states in $^{88}$Sr and $^{86}$Sr
can be well described by a simple model that only includes the $^3\Pi_u$
($0_u^+$) curve. Consequently, significant mixing between $^3\Pi_u$ and
$^1\Sigma_u^+$ must occur for bound states deeper in the well. Perturbation
theory dictates that the size of the $^1\Sigma_u^+$ contribution to an otherwise
$^3\Pi_u$ bound state increases in the vicinity of a $^1\Sigma_u^+$ bound
state. 

A comparison of the bound state energies predicted by our adiabatic (single
channel) and perturbed (two channel mass scaled) model is shown in
Table~\ref{tab:testing} for isotopes $^{84}$Sr and $^{88}$Sr. For both isotopes
we have adjusted the wavefunction phases so that the energies of the $\nu=-3$
bound state energies match. This way we ensure a fair comparison of vibrational
splittings. In our model the perturbing short range $^1\Sigma_u^+$ bound state
crosses the $^1S_0$$+$$^3P_1$ dissociation limit when the reduced mass $2\mu
\approx 84$~u. Consequently, there are significant differences in the bound
state energies for $\nu=-5$ and deeper states. On the other hand, in the case of
$^{88}$Sr, the differences between the models are about an order of magnitude
smaller. This shows that it will be possible to experimentally determine which
of the isotopes has its $^1S_0$$+$$^3P_1$ asymptote perturbed. In our model
$^{84}$Sr is the perturbed isotope, but given our limited knowledge of the
strontium PA spectra, $^{86}$Sr could be the perturbed isotope as well. The
subtle discrepancies between the $^{86}$Sr experimental data and our theoretical
model from Section \ref{sec:longrange} might hint toward this hypothesis. This
uncertainty can be resolved experimentally by simply measuring the bound state
energies deeper in the potential well. One of the isotopes should clearly have
its vibrational spacings incompatible with  an interaction model that only
includes the channels belonging to the $^1S_0$+$^3P_1$ asymptote.

\subsection{Mass scaling of $1_u$ energy levels and the three channel model
\label{sec:massscaling3}}

The last step in the construction of our mass scaled model is the introduction
of the $1_u$ molecular state. The interaction potential matrix, as expressed in
the $\left\{0_u^+(^3\Pi_u), 0_u^+(^1\Sigma_u^+), 1_u(^3\Sigma_u^+,
^3\Pi_u)\right \}$ Hund's case (c) basis is now
\begin{equation}
	V_{\rm int} = \left(
		\begin{array}{c c c}
						\begin{array}{c}\mathcal{V}(^3\Pi_u;R)-\\-C_{3,0}R^{-3} \end{array}&
						 \sigma \xi_1(R) & 0\\
						\sigma \xi_1(R) &
						\begin{array}{c}
						 \mathcal{V}(^1\Sigma_u^+;R) + \\ + \Delta E \end{array}& 0\\		
						 0 & 0 & \begin{array}{c}\frac{1}{2}\big(\mathcal{V}(^3\Pi_u; R)+\\ +\mathcal{V}(^3\Sigma^+_u; R)\big)- \\-{C_{3,1}}/{R^{3}}
						 \end{array} \\	
		\end{array}
	\right)\,.
\end{equation}
Similarly, the rotational Hamiltonian
\begin{equation}
	V_{\rm rot} = \left(
		\begin{array}{c c c}
			J(J+1)+2 & 0 & -\sqrt{4J(J+1)} \\
			0& J(J+1) & 0\\
			-\sqrt{4J(J+1)} & 0 & J(J+1) \\
		\end{array}
	\right)\,
\end{equation}
 now contains the Coriolis mixing terms like the long range interaction model
 from Section \ref{sec:longrange}. The kinetic Hamiltonian $T$ is again diagonal
 and the energy levels are calculated by solving the set of coupled
 Schr\"odinger equations $H\Psi = E\Psi$ for $H = T+V_{\rm int}+V_{\rm rot}$ and
 the three channel wavefunction $\Psi$.

 At this point reaching a mass scaled model for both $0_u^+$ and $1_u$ energy
 levels is relatively straightforward. Since the Coriolis mixing is a small
 correction to the $0_u^+$ energies we can hope that the mass scaling model
 utilizing the $^3\Pi_u$-$^1\Sigma_u^+$ curve crossing will stand except for
 cosmetic corrections to its parameters.

 Experimental $1_u$ energy level data is, to date, only available for the
 $^{88}$Sr and $^{86}$Sr isotopes. We can therefore apply the single channel
 mass scaling strategy as described in
 Section~\ref{sec:massscaling_singlechannel}, that is, fix the right number of
 bound states supported by the $1_u$ state. To this end, we have modified the
 short range $\alpha(^3\Sigma_u^+)$ parameter, which only affects the $1_u$
 quantum defect. Finally, we have run a least-squares optimization for both
 quantum defect $\alpha$ parameters and the curve crossing parameters $\Delta E$
 and $\sigma$ in order to obtain the final fit. All of the long range parameters
 ($C_{3,0}$, as well as $C_6$ and $C_8$ for both potential curves) remain the
 same.

A comparison of experimental energy levels for all strontium isotopes with those
predicted by our final three channel model is shown in
Table~\ref{tab:massscaling3}. Apart from the $^{84}$Sr $-1288$~MHz state
discussed at length in Section \ref{sec:vibrational} and not included in this
table, the theoretical energy levels match their experimental counterparts to
about 0.5~MHz on average. Mass scaling between states of $1_u$ symmetry is
acceptable, although we note that theoretical position of the $^{86}$Sr state at
$-159$~MHz is shifted by about 3~MHz. We have attempted to improve the accuracy
of this fit by changing the number of bound states supported by the $1_u$ curve,
but this resulted in a similar shift in the other direction.

For each bound state we also list channel
contributions $c_i$ calculated using $c_i = \int_0^{\infty}\left| \Psi_i(R)
\right|^2 dR$ and normalized via $c_1+c_2+c_3 = 1$. The channel contributions
again show the nonadiabatic effects described throughout this paper. For
instance, the Coriolis mixing between the $^1S_0$+$^3P_1$ $0_u^+$ and $1_u$ states
is clearly visible in the top $0_u^+$ bound states of $^{84}$Sr and $^{88}$Sr,
where the $1_u$ contribution reaches about $10~\%$. Similarly the pair of
$-8$~GHz states in $^{88}$Sr is again very strongly mixed with mutual
contributions as high as $14$~\%. On the other hand, the short range mixing
between the two $0_u^+$ states only significantly affects the $^{84}$Sr isotope
with contributions up to $3.9~\%$ for the $-228$~MHz state while for the
remaining isotopes it is below $0.04$~\% and as such is insignificant.

\begin{table}
\caption{Final $0_u^+$ and $1_u$ bound state energies as calculated from our
three channel mass scaled model described in Sec.~\ref{sec:massscaling3}.
Channel contributions are also given.\label{tab:massscaling3}}
\begin{ruledtabular}

\begin{tabular}{l r r r r r}
& \multicolumn{2}{c}{Energy levels (MHz)} & \multicolumn{3}{c}{Channel contributions (\%)} \\
\cline{2-3} \cline{4-6} \noalign{\vskip 2mm}
Isotope	& Experiment	& Theory	& $0_u^+ (^3\Pi_u)$	& $0_u^+ (^1\Sigma_u^+) $ &$1_u (^3\Sigma_u^+, ^3\Pi_u) $\\
\hline
$^{88}$Sr	& -0.435	& -0.426	& 90.991	& 0.000	& 9.009 \\
	& -23.932	& -23.856 	& 98.881	& 0.000	& 1.118 \\
	& -222.161	& -222.021	& 99.499	& 0.003	& 0.498 \\
	& -353.236	& -352.934	& 0.462		& 0.000	& 99.538 \\
	& -1084.093	& -1083.648	& 99.597	& 0.009	& 0.393 \\
	& -2683.722	& -2683.117	& 0.730		& 0.000	& 99.270 \\
	& -3463.280	& -3463.434	& 99.252	& 0.022	& 0.726 \\
	& -8200.163	& -8199.777	& 14.194	& 0.006	& 85.800 \\
	& -8429.650	& -8431.909	& 85.764	& 0.036	& 14.200 \\
	\hline \noalign{\vskip 2mm}
$^{86}$Sr	& -1.633	& -1.535	& 94.991	& 0.000	& 5.009 \\
	& -44.246	& -43.864	& 98.785	& 0.000	& 1.215 \\
	& -159.984	& -163.296	& 1.628		& 0.000	& 98.372 \\
	& -348.742	& -348.925	& 98.624	& 0.001	& 1.375 \\
	\hline  \noalign{\vskip 2mm}
$^{84}$Sr	& -0.320	& -0.341	& 89.470	& 0.025	& 10.505 \\
	& -23.010	& -23.541	& 89.183	& 0.548	& 10.269 \\
	&		& -32.855	& 10.125	& 0.074	& 89.800 \\
	& -228.380	& -228.259	& 95.130	& 3.912	& 0.957 \\
	&		&-1023.648	& 88.505	& 11.229& 0.266 \\
	&		&-1472.635	& 0.206		& 0.135	& 99.660 \\
\end{tabular}
\end{ruledtabular}
\end{table}

\section{Conclusion}

In conclusion we have measured the energies of four vibrational bound states of
the $^{86}$Sr molecule in an excited triplet state using photoassociation
spectroscopy of ultracold strontium atoms. The obtained data complements
previously reported photoassociation data for the remaining two bosonic isotopes
of strontium: $^{84}$Sr and $^{88}$Sr. We have provided an \emph{ab initio}
based theoretical model that correctly describes Coriolis mixing between $0_u^+$
and $1_u$ states. Only one of the previously reported energy levels, in
$^{84}$Sr, is in qualitative disagreement with our theory and we have suggested
that it may either be supported by a different potential curve, or be strongly
perturbed. We have used our theoretical model to provide Zeeman
$g$-factors for all of the considered photoassociation lines.

We have also shown, however, that a theoretical model that only takes into
account the two potential curves correlating to the $^1$S$_0$+$^3$P$_1$
asymptote will fail to correctly describe the positions of energy levels for all
isotopes via simple mass scaling, even despite being correct for one isotope at
a time. We attribute that to the short range avoided crossing between the
$^1$S$_0$+$^3$P$_1$ $^3\Pi_u$ curve and a higher excited state
$^1$S$_0$+$^1$D$_2$ $^1\Sigma_u^+$ potential: a short range $^1\Sigma_u^+$ bound
state would perturb the positions of the series of states observed in
photoassociation spectroscopy near the intercombination line. We have produced a
three channel model that take this mixing into account and is capable of
reproducing the available $0_u^+$ and $1_u$ energy levels. Finally, we suggest
that this theory can be tested by measuring more deeply bound vibrational
states. For at least one isotope the vibrational spacings should significantly
depart from those predicted by the long range model alone.

The results shown here will be important in the research concerning optical
Feshbach resonances near the intercombination line in Sr. The four resonances
reported in the paper can now be used to control the interactions between
$^{86}$Sr atoms in a similar manner as it was done in $^{88}$Sr \cite{Blatt2011,
Yan2013}. The mixing between states could also influence the possible use of Sr
as a means for measuring changes in the $m_p/m_e$ ratio \cite{DeMille2008,
Kotochigova2009}. The findings here also provide a template for
the description of mass scaling perturbed by short range avoided crossings in
other similar systems: a similar crossing between $^3\Pi_u$ and $^1\Sigma_u^+$
potentials has already been found in the calcium system~\cite{Allard2005}.

\acknowledgments

The authors wish to thank Piotr \.Zuchowski 
and Wojciech Skomorowski for their
help with setting up the calculations and useful discussions. We are also grateful to S.~G.~Porsev, M.~S.~Safronova and C.~W.~Clark for providing us with the results of their work before publication.
M. Y., B. J. D. and T. C. K. acknowledge support from the Welch 
Foundation (C-1579 and C-1669) and the National Science Foundation 
(PHY-1205946 and PHY-1205973). 
This work has been partially supported by the FNP TEAM Programme Project
\emph{Precise Optical Control and Metrology of Quantum
  Systems} nr TEAM/2010-6/3 cofinanced by the European Union Regional
  Development Fund and is part of the ongoing research program of National
  Laboratory FAMO in Toru\'n, Poland.

\appendix

\section{Fitting Photoassociation Spectra \label{sect:lineshape}}

\begin{figure}
\includegraphics[clip,width=.47\textwidth]{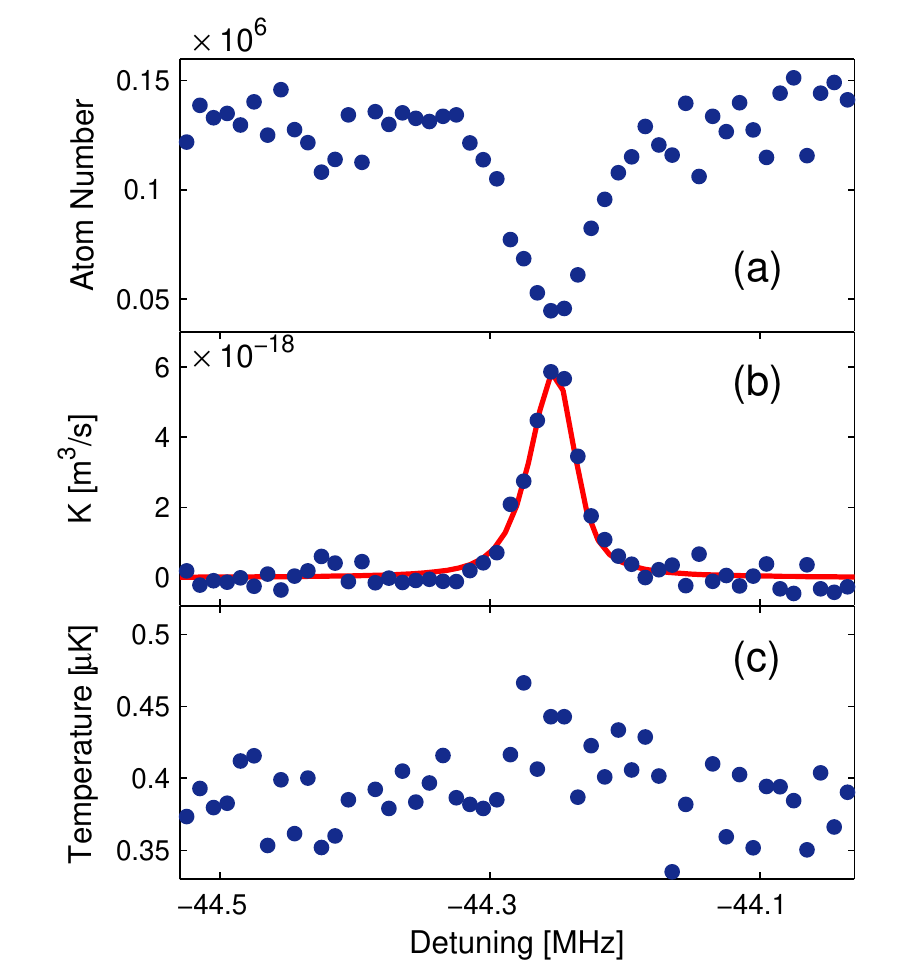}\\
\caption{(Color online) Spectroscopy of  the $0_u^+(\nu=-2)$ PA line of $^{86}$Sr.
(a) Atom number versus laser detuning from the one-photon $^1$S$_0-^3$P$_1$
atomic transition for  interaction time $t=59.4$\,ms.  (b) The effective
collision event rate constant derived from the atom loss using
Eq.\,\ref{KeffSimplifiedFurther}. (c) Temperature versus
  detuning shows little variation, supporting the assumption of constant sample
  temperature.
\label{-44MHzPASTime1936} }
\end{figure}

\begin{table}
\caption{Values of parameters of $^{86}$Sr PA lines extracted from
the experiment. Bound state energies are given for the $\nu=-1, -2,
-3$ levels of $0_u^+$ symmetry and the $\nu=-1$ level of $1_u$ symmetry, where $\nu$
counts levels down from the dissociation limit. These levels all have angular
momentum $J=1$. We also report the optical lengths $l_{\rm opt}$ for each of
the measured $0_{u}^+$ PA lines. The term $\eta^{\rm upper}$ describes the
additional molecular loss factor \cite{Theis2004,Blatt2011} and $\gamma^{\rm
upper}_{\rm laser}$ is the upper limit on the PA laser line width.}

	\begin{ruledtabular}
    \begin{tabular}{cccccc}
      $\nu$ &$^{86}$Sr & $\nu_0$ [MHz]
      	&$l_{\textrm{opt}}/I$[$a_\textrm{B}$/(W/cm$^2$)]&
      	$\eta^{\textrm{upper}}$
      	& $\gamma_{\textrm{laser}}^{\textrm{upper}}/2\pi$\\
      	& Series & & & & [kHz] \\
      \hline
      \,\,-1\,\,  & $0_u^+$ & -1.633(10)& $(3.8\sim6.5) \times 10^4$  &
      	1.7 & 10.5  \\
      -2  &   $0_u^+$ & -44.246(10)
      	& $(1.2\sim1.8) \times 10^4$  & 1.5 & 7.5 \\
      -1  &   $1_u$ & -159.984(50)
      & --       & --    & --  \\
      -3 & $0_u^+$ &\,\,-348.742(10)\,\, &
      $(5.7\sim10.3)\times 10^2$&1.8 & 12.0 \\
    \end{tabular}
    \label{table:86SrPAline}
	\end{ruledtabular}
\end{table}

Experiments determine the number of atoms in the trap after exposure to the PA
laser at frequency $f$ for interaction time $t$. Assuming constant sample
temperature, and loss described by $\dot{n}=-2Kn^2$, the time evolution of the
number of atoms is given by \begin{equation}\label{numberEvolution}
N(t)=\frac{N_0}{1+2 N_0 K t V_2 /V_1^2}, \end{equation} where $N_0$ is the
number of atoms without applying PA beams, $K$ is the effective collision event
rate constant, and $V_q$ ($q=1$ and 2) are the effective volumes defined by
\begin{equation}\label{eq:effectivevolumes} V_{q}=\int d^3\!r \,
e^{-qU(r)/k_{B}T}, \end{equation} with the trap potential $U(r)$, the Boltzmann
constant $k_B$, and the sample temperature $T$. For a high ratio of trap depth
to temperature, $\varepsilon_t/k_B T$, the effective volumes can be approximated
by \begin{eqnarray}\label{eq:AnalyticEffectiveVolumes} V_{q} = \biggr(\frac{2\pi
k_B T}{q m \overline{\omega}^2}\biggr)^{3/2}, \end{eqnarray} where $m$ is the
atomic mass, and $\overline{\omega}$ is the geometric mean of the angular
oscillator frequencies of the trap. Equations \ref{numberEvolution} and
\ref{eq:AnalyticEffectiveVolumes} yield
\begin{equation}\label{KeffSimplifiedFurther} K=\frac{4}{t}
\biggl(\frac{1}{N(t)}-\frac{1}{N_0}\biggl)\biggl(\frac{\pi k_B T}{m
\overline{\omega}^2}\biggl)^{3/2}, \end{equation} which allows direct
determination of spectra of $K$ from the atom-loss spectra.
Figure\,\ref{-44MHzPASTime1936} is an example near the $0_u^+(\nu\,=\,-2)$ PA line
of $^{86}$Sr.

The spectra of collision event rate constant $K$
(Fig.\,\ref{-44MHzPASTime1936}\,(b)) are fit using the formalism of Ciury{\l}o
\textit{et al.} \cite{Ciurylo2004}. This accounts for Doppler broadening and
photon recoil, which is necessary when the atomic temperature $T$ is lower than
the atomic recoil temperature $T_R=(h/\lambda)^2/(mk_B)=460$\,nK, for wavelength
of the PA laser $\lambda=689$\,nm. In our experiments, the sample temperature is
$T \sim 400$\,nK. In this regime, the collision event rate constant  is given by
\cite{Ciurylo2004} \begin{eqnarray}\label{KeffOriginal} K&=& \frac{k_B T}{h
Q_T}\int_{-\infty}^{+\infty} dy \,e^{-y^2} \int_{0}^{\infty} dx\,x \,e^{-x^2}
\mathcal{L}(f,y,x^2), \end{eqnarray} where $x$ and $y$ are dimensionless
variables, and 
\begin{equation}
	\label{L} \mathcal{L}(f,y,x^2) = {\eta
	\gamma_{\rm{mol}}{\gamma}_{s}/{(2\pi)^2}\over (f+y\Delta_D+x^2\Delta_T-E_{b}/h
	-E_{\rm{rec}}/h )^2 +(\eta'\gamma_{\rm{mol}}/ 4\pi)^2}. 
\end{equation}
with the thermal width $\Delta_T=k_BT/h$, the Doppler width
$\Delta_D=\sqrt{k_BT/m}/\lambda$, the natural linewidth of the excited molecular
level $\gamma_{\rm{mol}} = 2\pi \times 15$\,kHz, and the stimulated width
$\gamma_s = 2 (\sqrt{2\mu x^2 \Delta_T}/\hbar) \gamma_{\rm{mol}} \ell_{\rm{opt}}
\ll \gamma_{\rm{mol}}$, where $\ell_{\rm{opt}}$ is the optical length. Here, 
$Q_{T}=\left({2\pi k_{B}T \mu / h^2}\right) ^{3/2}$ is the partition function
for reduced mass $\mu=m/2$, $h$ is Planck's constant,  $E_b$ is the  PA line
center including the molecular ac Stark shift due to PA beams, and
$E_{\rm{rec}}=(h/\lambda)^2/(4m)$ is the photon recoil energy of an isolated
atom.  The parameter $\eta\geq1$ accounts for the extra molecular losses
observed in OFR experiments \cite{Theis2004,Blatt2011}, and $\eta^{\prime}
\gamma_{\rm{mol}}=\eta \gamma_{\rm{mol}} + \gamma_{\rm{laser}}$ with the line
width of the PA laser $\gamma_{\rm{laser}}$. We can only determine $E_{\rm{b}}$,
$\eta^{\prime}=\eta+\gamma_{\rm{laser}}/\gamma_{\rm{mol}}$, and $\eta
\ell_{\rm{opt}}$, and cannot independently determine $\eta$, $\ell_{\rm{opt}}$,
and $\gamma_{\rm{laser}}$. Truncation of the integral over collision kinetic
energy  has been neglected.

An example of the fitting  result is shown in
fig.\,\ref{-44MHzPASTime1936}\,(b). The  line center shifts linearly with the 
PA laser intensity, and we obtain spectra under a range of conditions and
extrapolate to zero intensity to obtain the unshifted  resonance position
$\nu_0$ (Table \ref{table:86SrPAline}). Fitting determines $\nu_0$ with a
precision of 5~kHz. There is additional systematic error in determining the
laser detuning with respect to atomic resonance  of 5 kHz, and we quote a  total
uncertainty of 10 kHz (Table\,\ref{table:86SrPAline}). The $\nu=-1$, $1_u$ binding
energy is quoted with increased uncertainty because the data showed  significant
temperature variation, complicating the analysis. Further details of the
experiment and the fitting procedure can be found in \cite{Yan2013PhD}.

By linearly fitting the values of $\ell_{\rm{opt}}$ at different $I$, we extract
$\ell_{\rm{opt}}/I$. 
The upper limit of $\eta$ ($\eta^{\rm{upper}}$) and the lower limit of
$\ell_{\rm{opt}}$ can be determined by fixing $\gamma_{\rm{laser}}=0$, while the
upper limit of $\gamma_{\rm{laser}}$ ($\gamma_{\rm{laser}}^{\rm{upper}}$,) and
the upper limit of $\ell_{\rm{opt}}$ can be determined by fixing $\eta=1$. 
Values of these parameters from the fitting are summarized in
Table\,\ref{table:86SrPAline}.

\bibliography{SrMassScaling}

\end{document}